\documentclass[pra,aps,twocolumn,showpacs,floatfix]{revtex4}
\usepackage{graphicx,psfrag}

\newcommand{\issmall}[2]{#1}

\begin{document}
\title{Spin textures in slowly rotating
Bose-Einstein Condensates}
\author{Erich J. Mueller}
\email{em256@cornell.edu}
\affiliation{
Laboratory of Atomic and Solid State Physics, Cornell University, Ithaca, New York 14853
}
\date{\today}

\begin{abstract}
Slowly rotating spin-1 Bose-Einstein condensates are studied through a variational approach based upon lowest Landau level calculus.  The author finds that in a gas with ferromagnetic interactions, such as $^{87}$Rb, angular momentum is predominantly carried by clusters of two different types of skyrmion textures in the spin-vector order parameter.  Conversely, in a gas with antiferromagnetic interactions, such as $^{23}$Na, angular momentum is carried by $\pi$-disclinations in the nematic order parameter which arises from spin fluctuations.  For experimentally relevant parameters, the cores of these $\pi$-disclinations are ferromagnetic, and can be imaged with polarized light.  
%All three of these textures are found to be ground states for some parameter range, a property shared by no other azimuthally symmetric textures.  From this analysis, the energy, angular momentum and magnetic moment are calculated as a function of the rotation speed and coupling constants.
\end{abstract} 
\pacs{03.75.Fi}
\maketitle

One of the most remarkable features of recent experiments on quantum degenerate alkali atoms has been the observation of quantized vortices in rotating Bose condensates of {\em spin polarized }$^{87}$Rb and $^{23}$Na \cite{vortexperiments}.  These vortices are a consequence of the irrotational nature of the superfluid flow in a {\em scalar} condensate (ie. one without any spin degrees of freedom).  If these same experiments were conducted in the absence of a magnetic field, the velocity of these spin-1 atoms would no longer be constrained to be irrotational; however, the curl of the velocity field would be fixed by spatial variations in the direction of the atomic spins \cite{spingauge}.  The vortices would therefore be replaced by intricate spin textures.  Here we predict the detailed spin patterns which will be found in such a rotating gas of spin-1 bosons.

Our first understanding of these structures came from Ho \cite{hospin} and Machida and Ohmi \cite{machidaohmi}, who independently proposed that spin textures replace vortices in spinor gases.    Subsequently several other authors investigated the detailed structure of these textures in both the slowly rotating \cite{slowrot} and fast rotating limits \cite{fastrot}.
The present work differs from these previous calculations in two vital ways: (i) it makes use of very simple variational wavefunctions from which the essential physics can be easily extracted, and (ii) it emphasizes the rotational properties of the local order parameter (which has both a vector and a nematic component.)  
The advantage of this approach, which
explicitly considers the order parameter symmetry, is 
that it
provides a scheme for classifying the spin textures, illustrates their structure, and most importantly, suggests a means for experimentally detecting them.

Although lacking our emphasis on symmetry arguments, similar theoretical techniques have recently appeared in an excellent preprint
by Reijnders et al. \cite{reijnders}.   That preprint surveys the properties of rotating spin-1 bosons for all possible parameters.  As such, it is complementary to the more focused study that we present.

Throughout this paper, we work within the mean-field lowest Landau level approximation.  This variational approach, detailed in section \ref{LLL}, is quantitatively accurate when the gas is so dilute that the interaction energy per particle is small compared to $\hbar \omega$, where $\hbar$ is Planck's constant, and $\omega$ is the frequency of the harmonic trap which confines the gas.  Our qualitative predictions, especially those based upon symmetry arguments, can be applied to denser systems.  Additionally, we restrict ourselves to the slowly rotating limit where the angular momentum per particle is on the order of a few $\hbar$.  In this limit, the angular momentum is carried by a small number of `elementary' textures, each of which is analogous to a single vortex.  We reserve the discussion of faster rotation speeds, where a regular lattice of these textures are found, to a future Article.

%As described in section~\ref{}, the spin-1 Bose gas is generically described by two interac
%We also restrict ourselves to values of the interaction parameters similar to those measured in experiments on Alkali atoms.  See \cite{reijnders} for a discussion of the wealth of possible states that are found if one considers more general interaction constants.

This Article is organized into the following topics: (i) a brief introductory example which illustrates the typical structure of spin textures; (ii) a review of the Hamiltonian and order parameter of a spin-1 gas of neutral Bose atoms; (iii) a description of our main theoretical approach; (iv) results; and (v) broader implications.  

\begin{figure}
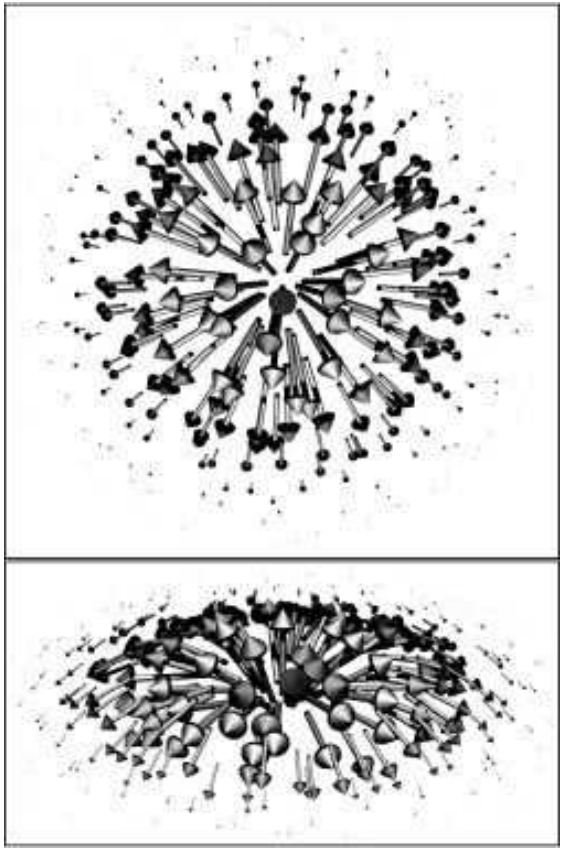

\issmall
{\includegraphics[width=210pt]{slowfig1.lr.epsf}}
{\includegraphics[width=210pt]{slowfig1.eps}}
\caption{Top view and perspective view of a $4\pi$ skyrmion.  
At the center (edge) the spins point out of (into) the page, while half way between the center and edge the spins lie in the plane of the page, rotating once as one circles the center.  The spins are located on a two dimensional plane, but point in three dimensions.
}\label{four}
\end{figure}

\section{Introduction}
In this section we discuss the two component Bose gas in order to develop
intuitive understanding of how internal degrees of freedom affect a rotating condensate. Later we will apply this intuition to the spin-1 gas.  In addition to simplicity, the two component gas  has the advantage of extensive experimental investigations.  As a concrete example one can consider two hyperfine states of $^{87}$Rb in a magnetic trap, as produced at JILA \cite{twocomponent}. The order parameter here is a two-component single-particle wavefunction ${\Psi}=(\psi_0,\psi_1)$, where $\psi_0$ and $\psi_1$ are the macroscopically occupied spatial wavefunction of each of the hyperfine states.  Using a phase-imprinting technique, the experimentalists are able to place one component, $\psi_1$ into a vortex state, while leaving $\psi_0$ in the ground state of the trap \cite{jilaphaseimprint}.  As a simple variational calculation shows \cite{var1}, this unusual `half-vortex' state is actually the ground state of the system for some rotation speed, and in the limit of weak interactions in two-dimensions can be written as
\begin{equation}
\Psi = \left(\begin{array}{c}\psi_0\\\psi_1\end{array}\right)=
 \left(\begin{array}{c}1\\z/d\end{array}\right)e^{-|z|^2/2d^2},
\end{equation}
where $z=x+i y$ is the coordinate in the plane, and $d=\sqrt{\hbar/m\omega}$ is the oscillator length, formed from Planck's constant $\hbar$, the particle mass $m$ and the frequency $\omega$ of the harmonic trap which is confining the particles.  
One can interpret $\Psi$ in terms of a pseudo-spin with polar angle $\theta$ and azimuthal angle $\phi$,  by writing
\begin{equation}
\Psi=\sqrt{\rho}e^{i\chi}
\left(\begin{array}{l}
\cos(\theta/2)\,e^{-i\phi/2}\\
\sin(\theta/2)\,e^{i\phi/2}
\end{array}
\right),
\end{equation}
where $\rho$ and $\chi$ represent the local density and phase.
The pseudo-spin behavior is then shown in figure~\ref{four}.  In particular, at the center of the cloud, the $\psi_1$ component vanishes, and the pseudo-spin points ``up."  At the edge of the cloud $\psi_0$ vanishes, and the pseudo-spin points ``down."  In between, the pseudo-spin rolls from up to down, covering $4\pi$ steradians of the sphere.  This texture is often referred to as a skyrmion in analogy to the work of T. H. Skrymi \cite{skyrmi}.
A detailed analysis of this two component Bose gas shows that these skyrmion textures are ubiquetous, and that in the rapidly rotating limit angular momentum is carried by a lattice of skyrmions \cite{twocomplat,uedalat}, where the geometry of the lattice is sensitive to the interaction parameters.

This (pseudo)spin-1/2 example teaches us that a rotating spinor condensate tends to carry angular momentum by twisting its spinor order parameter.   Leanhardt et al.  at MIT \cite{ketterleskyrmi} recently created the spin-1 and spin-2 analogy of this texture.  Similar structures have been seen in rotating superfluid $^3$He-A \cite{he3}, and in quantum hall systems \cite{bilayer}. The remainder of this Article predicts the equilibrium spin textures which will be found in rotating spin-1 condensates.

\section{Symmetries of the order parameter}\label{symord}
In this section we describe the order parameter of a spin-1 Bose gas.

As first discussed by  Ho \cite{hospin} and by Machida and Ohmi \cite{machidaohmi}, a trapped gas of spin-1 bosons interacting via a pairwise short-range interaction which is invarient under global spin rotations has a Hamiltonian of the form
\begin{eqnarray}\label{ham}
H&=&\sum_j\left(\frac{p_j^2}{2m}+U(r_j)\right)+\frac{1}{2}\sum_{ij}V_{ij},\\
V_{ij}&=&(c_0+c_2 {\bf S}_i\cdot {\bf S}_j)\delta(r_i-r_j),
\end{eqnarray}
where $i,j$ running from 1 to $N$ label the particles, $p_i$, $r_i$ and ${\bf S_i}$ are the momentum, position, and spin operators for each particle, $m$ is the particle mass, $U(r)=m\omega^2 r^2/2$ is the harmonic trapping potential, and $V$ is the pairwise interaction which is parameterized by two constants $c_0$ and $c_2$, which represent density and spin interactions.  The most commonly studied alkali atoms have $|c_2|/|c_0|\approx0.05\ll1$.  Ferromagnetic $(c_2<0)$ and antiferromagnetic $(c_2>0)$ interactions are respectively found in $^{87}$Rb and $^{23}$Na \cite{hospin}.
%The structure of spin textures in these two cases are very different, as is illustrated below.

In a Bose condensate of spin-1 atoms, the order parameter is the three component wavefunction $(\psi_1,\psi_0,\psi_{-1})$ representing the single-particle state which is macroscopically occupied.  The indices $1,0,-1$ represent the spin projection along the $\hat z$ direction.  As pointed out by Machida and Ohmi \cite{machidaohmi}, it is convenient to introduce a Cartesian representation of this wavefunction $\psi_x=(\psi_1-\psi_{-1})/\sqrt{2}, \psi_y=i (\psi_1+\psi_{-1})/\sqrt{2},\psi_z=\psi_0$, so that ${\bf \vec \psi}=(\psi_x,\psi_y,\psi_z)$ transforms as a vector under spin rotations.  
The order parameter is conventionally normalized so that the density is $\rho(r)=\vec\psi^\dagger(r)\cdot\vec\psi(r)$.
In the remainder of this paper we will freely move between the Cartesian [$\vec\psi=(\psi_x,\psi_y,\psi_z)$] and spherical [($\psi_1,\psi_0,\psi_{-1}$)] representation of the wavefunction, depending upon which is most useful.  In general, Latin indices (such as $a,b$) will refer to the Cartesian components, while Greek indices (such as $\mu,\nu$) refer to the spherical components.  Standard vector notation will be used when working within the Cartesian representation.

The local expectation value of the spin in the condensate is $\langle {\bf S}(r)\rangle=
\langle\sum_j \delta(r-r_j) {\bf S}_j\rangle = {\bf \vec \psi}^*(r)\times{\bf \vec\psi}(r)/i$.  
To minimize the spin interaction term for a
ferromagnetic/antiferromagnetic gas one maximizes/minimizes $|S|^2=|{\bf \vec\psi^*\times \vec\psi}|^2$.  Consequently a ferromagnetic gas prefers an order parameter with the structure 
\begin{equation}\label{ferstruc}
{\bf \vec\psi}={\bf n}+i{\bf m},
\end{equation}
 where $\bf n$ and $\bf m$ are real vectors with $\bf n\perp m$ and $|n|=|m|$, corresponding to the spin with magnitude $|S|^2=|\psi|^4$ pointing out of the plane in which $\bf n$ and $\bf m$ lie. 
When restricted to this maximally polarized sector,
the order parameter of the ferromagnetic gas can be taken to be the orthogonal triad $\bf n, m, S$, whose transformational properties are isomorphic to the group $SO(3)$.  Conversely, in the antiferromagnetic gas, the order parameter prefers to have $|S|^2=0$, which gives
\begin{equation}\label{antistruc}
{\bf \vec\psi}=e^{i\phi}{\bf n},
\end{equation} 
where $\bf n$ is a real vector and $\phi$ is a phase.  Thus antiferromagnetic interactions lead to an order parameter space isomorphic to $S_2\times U(1)/Z_2$, where $S_2$ is the sphere on which $\bf \hat n$ lies, $U(1)$ is the symmetry of the phase $\phi$, and the quotient with the two-element discrete group $Z_2$ represents the fact that $\bf \vec\psi$ is invariant under simultaneously reversing the direction of $\bf n$ and taking $\phi\to\phi+\pi$. Without the phase, the order parameter space is the group $S_2/Z_2$, which can be identified with the projective plane $RP_2$. The importance of this projective structure has been emphasized by Zhou \cite{zhou}.  

%It is amusing to note that the order parameter space of the antiferromagnetic gas covers $SO(3)$.  One produces this covering by noting that every rotation in $SO(3)$ corresponds to a rotation by an angle $\phi$ around some unit vector ${\bf \hat n}$, allowing us to map minimally polarized states, $\vec\psi=e^{i \phi}{\bf \hat n}$, onto the elements of $SO(3)$.  This mapping is not one-to-one, as in $SO(3)$ all rotations by the angle $\phi=0$ are equivalent, regardless of the direction $\hat n$.

A more intuitive approach to understanding the symmetries of the spin-1 condensate comes from considering the moments of the local spin operator.  A ferromagnetic interaction favors a maximally polarized state; for example $(\psi_1,\psi_0,\psi_{-1})=(1,0,0)$, for which the local spin is $\langle {\bf S}\rangle ={\bf \hat z}$.  In this case one can take $\langle {\bf S}\rangle$, an object which transforms as a vector, to be the relevant order parameter for discussing spin textures.  Antiferromagnetic interactions favor a minimally polarized state; for example $(\psi_1,\psi_0,\psi_{-1})=(0,1,0)$, for which $\langle {\bf S}\rangle={\bf 0}$.  To find a useful order parameter, one must consider the fluctuations of the spin $\delta S_a\delta S_b={\rm Re}\, \langle S_a S_b\rangle-\langle S_a\rangle\langle S_b\rangle$ \cite{whyreal}, which for the $(0,1,0)$ state is $\delta S_a \delta S_b=\delta_{ab}(1-\delta_{a z})$, where $\delta_{ab}$ is the Kroneker delta.  Consequently, this state has no spin fluctuations in the $\hat z$ directions, and isotropic fluctuations in the $x$-$y$ plane.  Due to the azimuthal symmetry, the spin fluctuation tensor $\delta S_a \delta S_b$ maps onto itself under rotation by $\pi$ radians.   Equivalently, one can say that the spin fluctuations represent a nematic order parameter (meaning that it transforms under rotation as an arrow with no head on it.)

As a consequence of the disparate types of spin textures which involve vector and nematic order parameters, gases with ferromagnetic and antiferromagnetic interactions behave in quite distinct manners.

\subsection{Local Vorticity}\label{localvort}
We now discuss how the symmetries of the order parameter are related to the curl of the velocity field,
\begin{equation}
{\bf v_s} = \frac{\hbar}{2 i m}\left(\psi_a^* \nabla \psi_a-\psi_a \nabla \psi^*_a\right),
\end{equation}
where repeated indices are summed over.
The key observation is that
unlike a scalar condensate, the spin-1 condensate can have a velocity field with extended vorticity.  This property is most simply understood by introducing a tensor
\begin{equation}
Q_{ab}=\frac{\psi_a^* \psi_b}{\sum_c \psi_c^*\psi_c},
\end{equation}
which carries all information about the local spin order parameters but no information about the density or phase.  By construction Q is Hermitian ($Q_{ab}=Q_{ba}^*$) and has trace $1$.  We decompose $Q$ into a sum of irreducible tensor operators,
\begin{eqnarray}
Q_{ab}&=& Q^{(0)}\delta_{ab}/3+i \epsilon_{abc} Q^{(1)}_c/2+Q^{(2)}_{ab},\\
Q^{(0)}&=& 1,\\
{\bf Q^{(1)}}&=& {\bf s}={\langle \bf S\rangle}/\rho,\\
Q^{(2)}_{ab} &=&(2/3) \delta_{ab}-(\langle S_a S_b\rangle+\langle S_b S_a\rangle)/2\rho^2,
\end{eqnarray} 
where $\epsilon_{abc}$ is the totally antisymmetric unit tensor and $\rho$ is the density.
The three irreducible components are a constant scalar $Q^{(0)}$, a vector $\bf Q^{(1)}$ which represents the local spin order, and a symmetric traceless tensor $Q^{(2)}$ which represents the local spin fluctuations.  The magnitude of these various components are constrained by the fact that $Q$ is idempotent ($Q^2=Q$), and hence
\begin{eqnarray}
0&=&{\rm Tr }(Q^2)-({\rm Tr} Q)\\
&=&-2/3+{\bf Q^{(1)}\cdot Q^{(1)}}/2+{\rm Tr} ((Q^{(2)})^2).
\end{eqnarray}
The last two terms are non-negative scalars which represent the amount of vector and nematic order.  This expression clearly shows that these are competing order parameters, as increasing one of them requires reducing the other.

As demonstrated in appendix~\ref{merminho}, the vorticity can be written in terms of $Q$ as
\begin{eqnarray}\label{mho}
{\bf \nabla\times v_s} &=& i\frac{\hbar}{m} Q_{ab} (\nabla Q_{bc}) \times (\nabla Q_{ca}).
\end{eqnarray}
In the spin polarized case $2 Q_{ab}=\delta_{ab}-s_a s_b + i \epsilon_{abc} s_c$ (with ${\bf s\cdot s}=1$) one recovers the Mermin-Ho \cite{mermho} relationship
\begin{equation}
{\bf \nabla\times v_s} = \frac{\hbar}{m} \epsilon_{abc}\, s_a (\nabla s_b)\times (\nabla s_c).
\end{equation}
In the purely (uniaxial) nematic case $Q_{ab}=n_a n_b$ (with ${\bf n\cdot n}=1$), one instead finds an irrotational flow, ${\bf \nabla\times v_s=0}$. Note that equation (\ref{mho}) is correct for all spin, not just spin 1.

In a more geometric language, the superfluid velocity 
is the phase one-form associated with ${\bf \psi}$, its curl is the phase two-form, and $Q$ is the operator which projects onto ${\bf \psi}$.  
The relationship (\ref{mho}) between the projector and the phase two-form, is generally valid, and shows up in other contexts \cite{twoform}.

\subsection{Visualization}
The ferromagnetic gas has a vector order parameter $\langle {\bf S}\rangle$ which is visualized by drawing a vector at each point in space (as in figure \ref{four}).  
The antiferromagnetic gas has a more complicated order parameter.  
In the strongly antiferromagnetic limit, the order parameter has the form of eq.~(\ref{antistruc}), and
one can visualize the order by plotting a rod aligned along the direction $\bf \hat n$ at each point in space.  In general, the order parameter deviates from eq.~(\ref{antistruc}), and it is more convenient to describe the order in terms of $Q^{(2)}$ or $Q^{(s)}_{ab}=(Q_{ab}+Q_{ba})/2=(1/3) \delta_{ab}+Q^{(2)}_{ab}$.  The tensor $Q^{(s)}$ is simpler to work with than $Q^{(2)}$ as it is positive semidefinite (meaning that all eigenvalues are greater or equal to zero).  In the case where $\psi$ is given by (\ref{antistruc}), then $Q^{(s)}$ has a single non-zero eigenvalue whose eigenvector points in the $\hat n$ direction.  In the general case, one can define $\hat n$ to coincide with the direction of the largest eigenvector of $Q^{(s)}$.

Generically $Q^{(s)}$ has three distinct eigenvalues, and therefore describes a {\em biaxial nematic}, which cannot be represented solely in terms of $\bf \hat n$.
For a more complete picture of the local nematic order, one replaces the rods by elipsoids whose three principle axes coincide with the eigenvectors of $Q^{(s)}$. The length of the principle axes are taken to be proportional the eigenvalues.  These ellipsoids are easily constructed by treating $Q^{(s)}$ as a linear transformation to a sphere.  Starting from a sphere $S$, the resulting ellipsoid is defined as the set of points satisfying
$r^\prime_a=Q^{(s)}_{ab} r_b$, for ${\bf r}\in S$.   In the uniaxial limit, $Q_{ab}^{(s)}=Q_{ab}=n_a n_b$, two of the ellipsoid's axes have length zero, and this degenerate ellipsoid becomes a rod pointing in the direction $\bf \hat n$.  In the ferromagnetic limit $2Q_{ab}^{(s)}=\delta_{ab}-s_a s_b$, the ellipsoid becomes a disc, whose normal is the spin vector $\bf s$.

In both the ferromagnetic and antiferromagnetic gas, the spins and directors point in three dimensional space.  Due to this intrinsic three dimensionality, it is impossible to faithfully represent the spin textures on a two-dimensional sheet of paper.  Consequently, in this Article we only provide sketches of the simplest textures.  Animated representations of the more complicated structures can be found in the EPAPS archive \cite{epaps}.

For simplicity, wherever possible we attempt to describe the nematic textures solely in terms of $\bf \hat n$.

\section{Lowest Landau level calculus}\label{LLL}
In this section we present a calculational scheme for studying Eq.~(\ref{ham}).  Our approximations are motivated by the limit where interactions are weak compared to the trapping potential.  Weak interactions are naturally reached under fast rotation,
 as in the rotating frame the centrifugal force effectively reduces the trap strength, causing the cloud to spread out over a larger area, reducing the density and hence the interactions.  Although motivated by this ultra-dilute limit, our approach gives an effective approximation to the properties of a system with much stronger interactions.  For example, both static \cite{rokhsar} and dynamic \cite{vortmotion} properties of vortex lattices in a scalar condensate have been very successfully modeled by this method.

In the rotating frame the Hamiltonian is shifted to $H^\prime=H-{\bf \Omega\cdot L}$, where
the vector $\bf \Omega$ points along the axis of rotation with modulus given by the rate of rotation, and $\bf L$ is the angular momentum vector of the system.  We restrict ourselves to two dimensions (taken to be the x-y plane), and rotate about $\hat z$ axis.  This dimensional restriction is consistent with the weakly interacting limit, where only a single mode of the trapping potential is occupied in the $\hat z$ direction.  To extend this method to describe the more general (3D) situation, one can use a Thomas-Fermi approximation as in \cite{homanyvortices}.

In the absence of interactions, the eigenstates of the two-dimensional single particle Hamiltonian $H^\prime_{\rm sp} = p^2/2m + m\omega^2 r^2/2 -\Omega (r_x p_y-r_y p_x)$ have energies $E_{nm}= \hbar\omega+\hbar(\omega+\Omega)n+\hbar(\omega-\Omega)m,$ and angular momentum $L_{nm}=\hbar(n-m),$ with $n,m=0,1,\ldots$  If the interaction energy per particle is sufficiently small compared to $\hbar(\omega+\Omega)$ then one can build up the quantum state from the ``Lowest Landau Level" (LLL), made up of the states with $n=0$.  Here we restrict our analysis to the mean field level, where one single particle state is macroscopically occupied.  The most general single particle state in the LLL can be written as
\begin{equation}\label{llwf}
\psi_\mu(x,y)=\sum_j c_{j\mu} \frac{ z^j}{\sqrt{\pi j!}} e^{-| z|^2/2},
\end{equation}
where $ z=(x+i y)/d$ is the coordinate in the plane in complex notation, scaled by the oscillator length  $d=\sqrt{\hbar/m\omega}$.  
The coefficients $c_{j\mu}$ can also be expressed in Cartesian notation ${\bf c}_j=(c_{jx},c_{jy},c_{jz})$.
We normalize this wavefunction so that
\begin{equation}
\int\!\!dx\,dy\,\,{\bf \vec \psi^*\cdot\vec\psi}
=\sum_j {\bf c_j^*\cdot c_j}=1.
\end{equation}
Substituting this ansatz into (\ref{ham}), and neglecting the zero point energy, the scaled energy per particle in the rotating frame becomes
\begin{widetext}
\begin{eqnarray}
{\cal E}&=&
\frac{\langle H-\Omega L_z\rangle}{N\hbar(\omega-\Omega)}\\\nonumber
&=&
\frac{1}{\hbar(\omega-\Omega)}\int\!\!dx\,dy\,\left(\frac{\nabla \vec\psi^*\cdot\nabla\vec\psi}{2 m}
+\frac{m\omega^2 r^2}{2} (\vec\psi^*\cdot\vec\psi)
- \hbar \Omega (\vec\psi^*(x\partial_y-y\partial_x)\vec\psi)
+\frac{c_0 N}{2}({\bf \vec\psi^*\cdot\vec\psi})^2+\frac{c_2 N}{2}
|{\bf \vec\psi^*\times\vec\psi}|^2
\right)\\\label{var1}
&=&\sum_j \left[(j+1) {\bf c_j^*\cdot c_j} +\frac{\eta}{2}\left[(1+\bar c_2)({\rm Tr}\,{\bf d_j^*\cdot d_j)}-\bar c_2 |{\rm Tr}\,{\bf d_j}|^2\right]\right],\\
(d_j)_{st} &=& 2^{-j}\sum_k {{j}\choose{k}}^{1/2} (c_{j-k})_s (c_k)_t,
\qquad\qquad
\eta=\frac{N c_0}{2\pi(\omega-\Omega)}.\label{etadef}
\end{eqnarray}
\end{widetext}
Here `${\rm Tr}$' signifies taking the trace of a matrix.
The simplicity of these equations illustrate the
utility of the Cartesian representation of $\psi$.
In the LLL, the system is parameterized by two dimensionless couplings; $\eta$, which measures the ratio between the interaction strength and the level spacing, and $\bar c_2=c_2/c_0$, the relative strength of spin and density interactions.  Larger $\eta$ corresponds to faster rotation.
If one rotates at speeds where $\Omega>\omega$, the centrifugal force becomes stronger than the trapping force and the system becomes unstable.  This centrifugal limit coincides with $\eta\to\infty$.  One can experimentally tune $\eta$ between $0$ and $+\infty$.

In this Article we find the ground state as a function of $\eta$ and $\bar c_2$ by minimizing (\ref{var1}) with respect to the parameters ${\bf c_j}$ using a conjugate gradient method.  We truncate the allowed values of $j$ to be $j=1,2,\cdots,J$.  Convergence is ensured by taking several values of $J$, ranging from $2$ to $128$, and finding that for sufficiently large $J$ the energy saturates.  Although this variational approach is much simpler than directly minimizing the Gross-Pitaevskii equation, it is still a nontrivial numerical task.  Since ${\bf c_j}$ is a three component complex object, one has the equivalent of $6 J$ real variational parameters.  Symmetries can be used to reduce the number of independent parameters.
%  As is discussed below, the number of parameters can be greatly reduced by making extensive use of symmetries.  We emphasize, however, that all of our results are verified by unconstrained minimization of (\ref{var1}).  
All our results use the approximate physical value, $\bar c_2=\pm0.05$ \cite{physicalc2}.  For a discussion of other values of $\bar c_2$, see \cite{reijnders}.

Not surprisingly, the energy landscape is quite complicated with many metastable local minima.  Computationally, this large degree of metastability makes finding the absolute ground state quite difficult.  In section~\ref{symsec} we show how symmetries can be used to help find these minima.  This symmetry approach is augmented by 
a brute-force search strategy where we
run our minimization routines from a very large number of initial conditions, and select out the lowest energy minima found.  Experimentally, the presence of so many local minima implies that the state found in the laboratory will depend strongly upon the method of preparation.  Moreover, one would expect to see domain structures where different parts of the sample exhibit different phases.

\subsection{A More General Ansatz}
Although all the numerical calculations presented here involve equation (\ref{var1}), it should be noted that 
by introducing an extra variational parameter, one can greatly extend the realm of validity of the LLL approximation.  The resulting equations are not much more complicated than (\ref{var1}). %however the solutions depend on a third dimensionless parameter, $\omega/(\omega-\Omega).$

The extended LLL ansatz scales the lengths in (\ref{llwf}) by a new variational parameter $\nu$ so that $z=\nu^{1/2}(x+I y)/d$.  Neglecting zero point motion, the energy per particle is
\begin{eqnarray}
\frac{\langle H^\prime\rangle}{N\hbar}
&=&\left[\frac{\omega}{2} \left(\nu+\frac{1}{\nu}\right)-\Omega\right] I_1
+ (\omega-\Omega)\frac{\eta}{2} \nu I_2,\\
I_1 &=&\sum_j (j+1) {\bf c_j^*\cdot c_j},\\
I_2 &=&\sum_j \left[(1+\bar c_2)({\rm Tr}\,{\bf d_j^*\cdot d_j)}-\bar c_2 |{\rm Tr}\,{\bf d_j}|^2\right].
\end{eqnarray}
Minimizing with respect to $\nu$ with $\{{\bf c_j}\}$ fixed yields
\begin{eqnarray}\label{var2}
H^\prime&=&I_1\hbar\left[\omega\sqrt{1+\eta\frac{I_2}{I_1}\frac{\omega-\Omega}{\omega}}-
\Omega\right],\\
\nu^{-2}&=&1+\frac{\eta}{2}\frac{I_2}{I_1}\frac{\omega-\Omega}{\omega}.
\end{eqnarray}
In the limit $\lambda=\eta [(\omega-\Omega)/\omega][I_2/I_1]\ll 1$ one recovers (\ref{var1}).  One can minimize (\ref{var2}) by the methods already discussed.  Note that instead of depending upon only two experimental inputs, $\eta,$ and $\bar c_2$, the minima in the extended LLL ansaztz also depends on the dimensionless ratio $(\omega-\Omega)/\omega$.

\begin{figure}
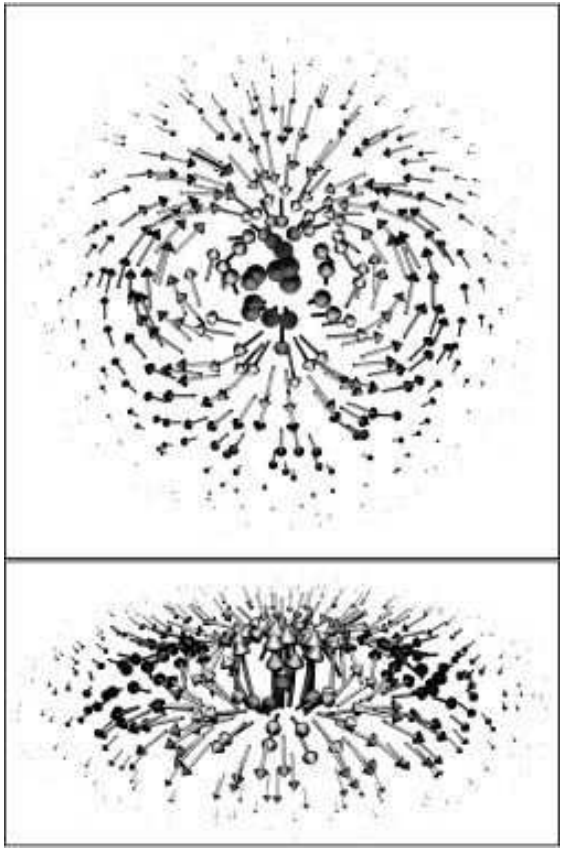

\issmall
{\includegraphics[width=210pt]{slowfig2.lr.epsf}}
{\includegraphics[width=210pt]{slowfig2.eps}}
\caption{Top view and perspective view of an $8\pi$ skyrmion.  This texture differs from figure~\ref{four} in that the planar projection of the spins rotates twice as one circles the origin.}\label{eight}
\end{figure}

\section{Symmetries of the spin textures}\label{symsec}
\subsection{Symmetries}
Given the complexity of the energy landscape,  an unrestricted minimization of eq.~(\ref{var1}) is a daunting task.  The large number of variational parameters makes the computation expensive, and the large number of metastable minima means that an exhaustive search is needed to find the absolute minimum.  Moreover, a classification scheme is needed to describe the minimal configurations once they are found.
We solve all of these problems by studying the possible symmetries of the spin textures.  In section~\ref{observe}, we discuss how these symmetries can be used to detect the textures.  This strategy was inspired by discussions with Dan Rokhsar, who, with Dan Butts, used similar techniques in understanding vortex lattices in scalar condensates \cite{rokhsar}.

Our Hamiltonian (\ref{ham}) is separately invarient under each of the following operations: (i) rotating spatial coordinates about the origin by an arbitrary angle $\theta$, 
\begin{eqnarray}
{\bf \psi}(r_a) &\longrightarrow& {\bf R}_{ab}(\theta) {\bf \psi}(r_b)\\
{\bf R}(\theta) &=& \left( 
\begin{array}{rl}
\cos\theta&\sin\theta\\
-\sin\theta&\cos\theta
\end{array}
\right);
\end{eqnarray}
(ii) Rotating all spins by arbitrary Euler angles $\phi$, $\theta$, and $\chi$, 
\begin{eqnarray}
\psi_\mu&\longrightarrow&{\cal R}_{\mu\nu}(\theta,\phi,\chi)\psi_\nu\\
{\cal R}(\theta,\phi,\chi)&=&
e^{-i \chi S_z}
e^{-i \theta S_x}
e^{-i\phi S_z};
\end{eqnarray}
where $S_x,S_y, S_z,$ are the operators for the spin components; 
(iii) simultaneously reflecting spatial coordinates across the line perpendicular to the unit vector $\bf \hat n$, and applying the time reversal operator (which takes $\psi_\mu\rightarrow \psi_{-\mu}^*$), 
\begin{eqnarray}
\psi_\mu({\bf r})&\longrightarrow&\psi_{-\mu}^*({\bf\bar R_{\hat n} r})\\
{\bf \bar R_{\hat n}} {\bf r} &=&
{\bf r}-2{\bf \hat n\cdot r};\end{eqnarray}
(iv) reflecting the spin across the plane perpendicular to the unit vector $\bf\hat n$,
\begin{eqnarray}
\vec\psi &\longrightarrow& \bar{\cal R}_{\hat n} \vec\psi\\
\bar{\cal R}_{\hat n} \vec\psi &=& \vec\psi-2{\bf \hat n\cdot\vec\psi}.
\end{eqnarray}
and (v) a global gauge transformation by an arbitrary phase $\chi$,
\begin{eqnarray}\label{gauge}
\psi_\mu&\longrightarrow & e^{i \chi}\psi_\mu.
\end{eqnarray} 

Although the Hamiltonian is invarient under each of these transformations, the mean-field wavefunction is not.  We will classify spin textures by the way in which they break these symmetries.

\subsection{Scalar}
To see how these symmetries manifest in spin textures, it is helpful to first consider a scalar condensate \cite{rokhsar}, for which only symmetries (i), (iii), and (v) are relevant.  In the lowest Landau level approximation, the non-rotating scalar ground state is $\psi(r)=e^{-|r|^2}$, which is invarient under (i) and (iii), but not (v): i.e. the only broken symmetry in the non-rotating condensate is the gauge symmetry.

At appropriate rotation speeds the ground state contains a single vortex, and (in the lowest Landau level) the order parameter becomes $\psi=z e^{-|r|^2/2},$ where, as usual, $z=x+i y$.  This single vortex state is no longer invarient under any of the transformations, (i), (iii), or (v).  However, it is invarient under appropriate combinations.  For example, rotating the spatial coordinate by any angle $\theta$ transforms $z\to e^{i\theta} z$, and can therefore be undone by a global gauge transformation (\ref{gauge}) with $\chi=\theta$.  This is an example of a continuous symmetry, as for each $\theta$ there is a combined spatial rotation and gauge transformation which leaves the state invarient.  This symmetry is not restricted to the lowest Landau level approximation, and it should be clear that any condensate with a single vortex at the center will have this symmetry, expressed as
\begin{equation}
\psi({\bf R}(\theta){\bf r}) = e^{i m\theta}\psi({\bf r}),
\end{equation}
where $m$ is a fixed integer which gives the number of quanta of circulation which are concentrated in the vortex core.  An ordinary vortex has $m=\pm1.$

At faster rotation speeds, the condensate will contain a small cluster of vortices.  As an example, one can imagine two vortices, symmetrically placed at $z=\mp z_0$, so that the order parameter is
$\psi = (z-z_0)(z+z_0)e^{-|r|^2/2}$.   No continuous symmetry exists here.  However,  there does exist a discrete symmetry in that this state is invarient under rotating space by $\theta=\pi$.  Similarly, three vortices which form an equilateral triangle will be invarient under a rotation by $\theta=2\pi/3$.  Such discrete symmetries are generally of the form
\begin{equation}\label{discrete}
\psi({\bf R}(2\pi/a){\bf r})=e^{2\pi i m/a} \psi({\bf r}),
\end{equation}
where $a$ is an integer which, together with the integer $m$, describes the symmetry of the cluster of vortices. 

The wave function for a cluster of vortices with a reflection plane will be invarient under a combination of a reflection across a fixed line, time reversal (complex conjugation), and a gauge transformation,
\begin{equation}
\psi(\bar{\bf R}_{\hat n}r)^* = \pm \psi({\bf r})
\end{equation}
where the line of reflection is perpendicular to the unit vector $\bf \hat n$.  We have written the gauge transformation as a multiplication by $\pm1$, as the only consistent phases are $\chi=\pi,0$.

\subsection{Spin 1}
As with vortices in a scalar gas, spin textures in a spinor condensate are characterized by their properites under different combinations of symmetry operations.  The spin degrees of freedom only introduce additional symmetries.
%We have seen that the vortex structure in a rotating scalar gas are characterized by their symmetries under a combination of spatial rotations/reflections and gauge transformations.  The spin textures in a spinor condensate will be similarly characterized, except that one also has to consider spin rotations/reflections.

%As with the scalar case, a
At low rotation speeds one expects to find a continuous symmetry, generically involving the simultateous rotation of space, spin, and phase.  Without any loss of generality, we can take the spins to rotate about the $\hat z$ axis,  in which case the symmetry is formally described by the statement
that  there exists constants $m$ and $n$ such that for all $\theta$,
\begin{equation}\label{rot1}
 \psi_\mu({\bf R}(\theta) r) = e^{i m \theta} {\cal R}_{\mu\nu}(n\theta) \psi_\nu(r),
\end{equation}
where ${\bf R}(\theta)$ represents a spatial rotation about the origin by an angle $\theta$,
${\cal R}_{\mu\nu}(\phi)$ represents a spin rotation about the $\hat z$ axis by an angle $\phi$, and $m,n$ count the number of times the phase and spin angles rotate in comparison to the spatial angle $\theta$.  Single valuedness of the left hand side of (\ref{rot1}) constrains $m$ and $n$ to either both be integers, or both be half-integers.  Further explanation of this constraint will be given in section~\ref{math}.

Textures obeying (\ref{rot1}) will be described as the `elemental,'  `single,' or `azimuthally symmetric' spin textures.  They are the building blocks for all more complicated strucures.

At higher rotation speeds, this continuous symmetry is replaced with a discrete one.  In analogy to 
the scalar case (\ref{discrete}), the discrete rotational symmetry is of the form 
\begin{equation}\label{rot2}
 \psi_\mu({\bf R}(2\pi/a) r) = e^{i m 2\pi/a} {\cal R}_{\mu\nu}(n 2\pi/a) \psi_\nu(r),
\end{equation}
where $a$ is a fixed integer describing the geometry of the texture.

In addition to rotational symmetries, the state may have reflection symmetries. The most general form of reflection symmetry which we consider consists of simultaneously performing transformations (iii), (iv), and (v), namely reflecting the spatial coordinates accross a line, time reversal, reflecting the spins accross a plane, and multiplying by the factor $\pm1$.   By globally rotating our spins, we can always assume that the spin and spatial reflection planes coincide.
 Mathematically this symmetry can be expressed as
\begin{equation}\label{reflect}
\left( \psi_{-\mu}({\bf \bar R_{\hat n}} r)\right)^* = \pm (\bar{\cal R}_{\hat n})_{\mu\nu} \psi_\nu(r),
\end{equation}
where $\hat n$ is the normal to the reflection plane, 
${\bf \bar R_{\hat n}}$ is the spatial reflection and $ (\bar{\cal R}_{\hat n})_{\mu\nu}$ is the spin reflection.  One can also consider the case where no spin reflection is performed
\begin{equation}\label{partreflect}
\left( \psi_{-\mu}({\bf \bar R_{\hat n}} r)\right)^* = \pm \psi_\nu(r).
\end{equation}
We will refer to (\ref{reflect}) as a reflection and 
(\ref{partreflect}) as a {\em partial reflection}.

\subsubsection{implications}\label{math}
Enforcing the various possible symmetries describe in equations (\ref{rot1}) through (\ref{partreflect}), greatly reduces the number of parameters which must be minimized.  Given the explict form of the rotation operators
\begin{eqnarray}
{\bf R}(\theta) z &=& e^{-i\theta} z\\
{\cal R}_{\mu\nu}(\theta) &=& \delta_{\mu\nu}e^{i\theta\mu},
\end{eqnarray}
and the lowest Landau level ansatz $\psi_\mu=\sum_j {c_{\mu j}}z^j e^{-|z|^2/2}$, the continuous symmetry (\ref{rot1}), restricts $c_{\mu j}$ to be zero unless
\begin{equation}
j= m+\nu n.
\end{equation}
Since $\nu=1,0,-1$, this requires any non-trivial texture to have $m,n$ both be integers, or both be half integers.  At most three $c_{\mu j}$'s are non-zero.  By combining an overall spin rotation, a spatial rotation, and a gauge transformation, these three parameters can be taken to be real.  Using the normalization constraint $\sum_{j,\mu} |c_{\mu j}|^2=1$, one is left with only two (real) variational parameters.

The discrete symmetries similarly reduce the number of parameters.  For example the discrete rotation symmetry forces $c_{\mu j}=0$ unless
\begin{equation}
j\equiv m+\nu n\quad ({\rm mod}\,\, a).
\end{equation}
Up to an overall phase, enforcing the reflection symmetry (\ref{reflect}) with ${\bf \hat n=\hat y}$, requires $c_{\mu j}$ is real.

\section{results}
\subsection{Azimuthally Symmetric Textures}
In the limit of very slow rotation, the angular momentum will be carried by a single `elemental' 
spin texture, 
characterized by the continuous symmetry in equation (\ref{rot1}).  This structure is
analogous to a single vortex.  At faster rotation speed a cluster of these textures will form.  Eventually the clusters will become regular lattices.  In this section the structure of the `elemental' spin textures will be explored.  
%A qualitative description of the results will be presented, followed by a variational calculation.  
%These conclusions have been verified by extensive numerics. 

\subsubsection{Ferromagnetic}\label{fersec}
The ferromagnetic gas displays
two different elemental textures, both of which are similar to the skyrmion encountered in the spin-1/2 case.  These two textures, illustrated in figures~\ref{four} and \ref{eight}, differ by the amount of solid angle ($4\pi$ or $8\pi$) traced out by the spin vectors as they move from pointing straight up at the center to straight down on the edge, and will therefore be referred to as $4\pi$ or $8\pi$ skyrmions.  If one looks down from above, as shown in the top panels of figures~\ref{four} and \ref{eight},   one can distinguish these textures by the number of times which the planar projection of the spin rotates when one circles the origin (once/twice for the $4\pi$/$8\pi$ texture).  Higher order textures (where angles greater than $8\pi$ are subtended) are never found to have continuous azimuthal symmetries, and are always more appropriately described as a collection of $4\pi$ and $8\pi$ skyrmions.   At faster rotation speeds, angular momentum is carried by a lattice of $4\pi$ or $8\pi$ skyrmions where their quantization axes takes on several different orientations.  These textures only cover a full $4\pi$ and $8\pi$ steradians of solid angle when they exist in isolation in an infinite system.  In a lattice, the skyrmions often touch before they can completely cover the sphere.

The $4\pi$ and $8\pi$  textures are described by (\ref{rot1}), with $m=n=1$ and $m=n=2$.  In a basis with spin projections (1,0,-1), the spinors take the form
\begin{eqnarray}\label{sym}
%\begin{array}{c}
|n=m=1\rangle &=& \left(\begin{array}{c}
1\\
a z\\
b z^2
\end{array}\right) e^{-|z|^2/2},\\
|n=m=2\rangle &=& \left(\begin{array}{c}
1\\
a z^2\\
b z^4
\end{array}\right)e^{-|z|^2/2}.
\end{eqnarray}
The real coefficients $a$ and $b$ are given by minimizing (\ref{var1}).

\begin{figure}
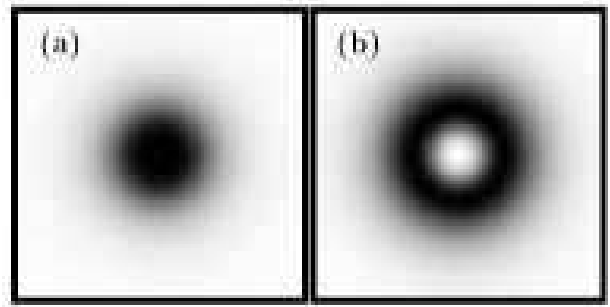

\issmall
{\includegraphics[width=234pt]{slowfig3.lr.epsf}}
{\includegraphics[width=234pt]{slowfig3.eps}}
\caption{Local vorticity $\hat z\cdot\nabla\times v$, for the $4\pi$ (a) and $8\pi$ (b) skyrmion.  Darker colors represent larger vorticity.}\label{skyrmcurl}
\end{figure}

In the $4\pi$ skyrmion, the velocity field has its greatest curl at the center of the texture, while for the $8\pi$ skyrmion the curl is greatest on a ring of finite radius (see figure~\ref{skyrmcurl}).  According to the Mermin-Ho relationship (\ref{mho}) one can attribute these two distinct behaviors to the fact that the spins `bend' fastest in these regions.

\begin{figure}
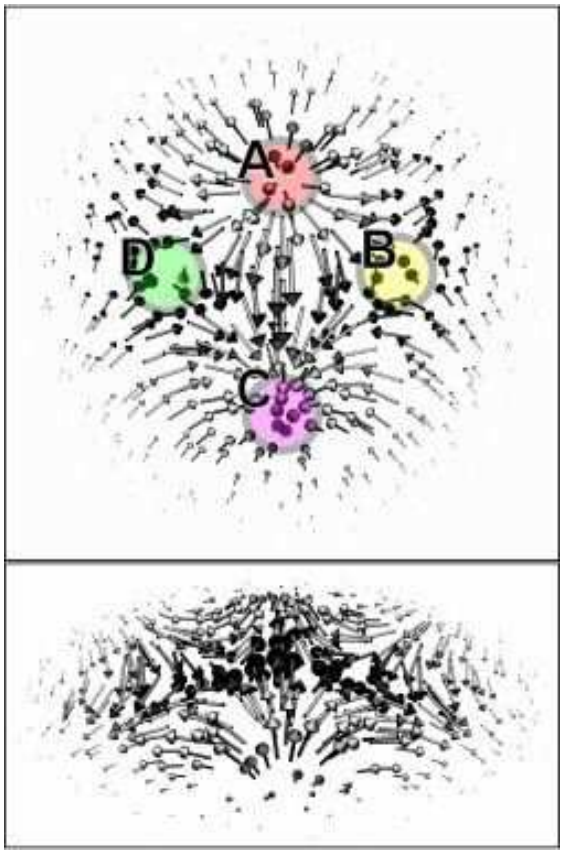

\issmall
{\includegraphics[width=210pt]{slowfig4.lr.epsf}}
{\includegraphics[width=210pt]{slowfig4.eps}}
\caption{(color online) Top view and perspective view of an $8\pi$ skyrmion.  This is the same texture as fig.~\ref{eight}, except all spins have been rotated by 90 degrees about the $\hat y$ axis.  This texture can be interpreted as a composite of four $4\pi$ skyrmions, whose cores have been marked with circles, and labeled by letters (A) through (D). The spins in (A) and (C) point out of the page and the planar projections of the spins wrap the equator of the order-parameter sphere in a positive sense, while (B) and (D) point into the page and the projected spins wrap the equator in a negative sense.  }
\label{doublerot}
\end{figure}

The $8\pi$ skyrmion can alternatively be interpreted as a composite of four $4\pi$ textures \cite{misnomer, fujita}.  This interpretation is illustrated in figure~\ref{doublerot}, where the spins in figure~\ref{eight} are rotated around the y-axis by 90 degrees.  There are four points where the spins are pointing into/out of the page.
These points lie on the outer edge of the ring of maximal $\nabla\times v$, and can be taken to be skyrmion cores.  Note that their positions are not unique, and by globally rotating all of the spins, these points move around in a ring.  This lack of uniqueness is a consequence of the high degree of symmetry of this state.  The $8\pi$ steradians of solid angle subtended by the texture can be divided evenly among these smaller skyrmions, allowing us to attribute $2\pi$ steradians to each of them.  Two of the sub-textures, labeled (A) and (C), have cores pointing out of the page, while the other two, labeled (B) and (D) point into the page.  Equidistance between these cores the spins lie in the x-y plane.  When circling (A) or (C) these planar spins rotate in a positive sense while around (B) or (D) they rotate in a negative sense.

\begin{figure*}
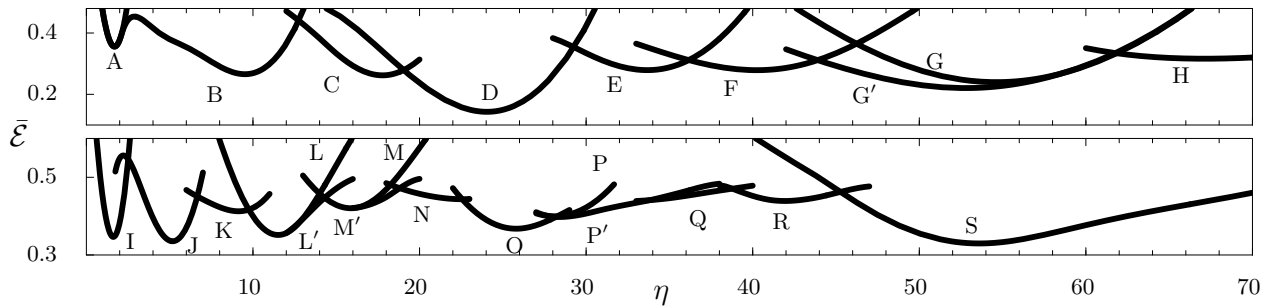

%\psfrag{en}[lb]{\large $\bar{\cal E}$}
%\psfrag{eta}{\large $\eta$}
%\psfrag{A}{\small A}
%\psfrag{B}[c]{\small B}
%\psfrag{C}[c]{\small C}
%\psfrag{D}{\small D}
%\psfrag{E}[c]{\small E}
%\psfrag{F}[c]{\small F}
%\psfrag{Gp}{\small G}
%\psfrag{G}[c]{\small G$^\prime$}
%\psfrag{H}[c]{\small H}
%\psfrag{I}{\small I}
%\psfrag{J}[rc]{\small J}
%\psfrag{K}[cl]{\small K}
%\psfrag{Lp}{\small L}
%\psfrag{L}{\small L$^\prime$}
%\psfrag{Mp}{\small M}
%\psfrag{M}[c]{\small M$^\prime$}
%\psfrag{N}[cr]{\small N}
%\psfrag{O}[c]{\small O}
%\psfrag{Pp}{\small P}
%\psfrag{P}[c]{\small P$^{\prime}$}
%\psfrag{Q}[t]{\small Q}
%\psfrag{R}[c]{\small R}
%\psfrag{S}{\small S}
%\psfrag{10}[c]{\small 10}
%\psfrag{20}[c]{\small 20}
%\psfrag{30}[c]{\small 30}
%\psfrag{40}[c]{\small 40}
%\psfrag{50}[c]{\small 50}
%\psfrag{60}[c]{\small 60}
%\psfrag{70}[c]{\small 70}
%\psfrag{0.3}[lb]{\hspace{-2mm}\small 0.3}
%\psfrag{0.4}[lb]{\hspace{-2mm}\small 0.4}
%\psfrag{0.5}[lb]{\hspace{-2mm}\small 0.5}
%\psfrag{0.2}[lb]{\hspace{-2mm}\small 0.2}
%\includegraphics[width=\textwidth]{tmp/wideenstrip.eps}\\
\issmall
{\includegraphics[width=\textwidth]{slowfigrenum7.lr.epsf}}
{\includegraphics[width=\textwidth]{slowfigrenum7.eps}}
\caption{(Wide)
Scaled energies $\bar {\cal E}$ as a function of rotation speed, parameterized by $\eta$ defined in Eq.~(\ref{etadef}).  Larger $\eta$ corresponds to faster rotation.  Top/bottom panels show ferromagnetic/antiferromagnetic interactions with $c_2=\mp0.05 c_0$.  Each curve represents a state of different symmetry, as described in the text and figures~\ref{struc} through \ref{struc2}. 
More detailed images of these states can be found in the EPAPS archive \cite{epaps}
Energy scalings (Top: $\bar {\cal E}={\cal E}^2-1.7125\eta$, Bottom: $\bar{\cal E}={\cal E}^2-1.8084\eta$) are chosen to aid in comparing these different curves.
}\label{strip}
\end{figure*}

The $4\pi$ skyrmion has lower energy than the uniformly polarized state $(\psi_1,\psi_0,\psi_{-1})=(1,0,0)$ if and only if $\eta>2/(1+c_2)$.  This is a second order (continuous) phase transition.  Between the region of stability of the $4\pi$ and $8\pi$ skyrmions, a more complicated texture is found.  Details of this intervening state will be given in section \ref{composite}.  A graphical comparison of the energies of these states as a function of $\eta$ is made in figure~\ref{strip}.

\subsubsection{Antiferromagnetic}\label{nemsec}
\begin{figure}
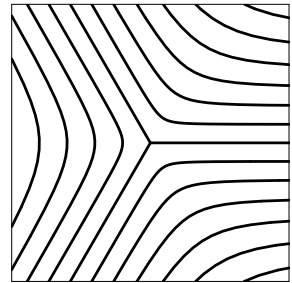

\issmall
{\includegraphics[width=105pt]{slowfig5.lr.epsf}}
{\includegraphics[width=105pt]{slowfig5.eps}}
\caption{Representations of a $\pi$ disclination in a nematic.  
Lines represent the local orientation of the nematic director.}
\label{pidisc}
\end{figure}

The antiferromagnetic gas, with its nematic order parameter, supports a much different set of textures than the ferromagnetic gas.  In particular, one finds that angular momentum is predominantly carried by $\pi$-disclinations, which as illustrated in figure~\ref{pidisc} are objects around which the nematic director rotates by 180 degrees ($\pi$ radians).  There is no consistent way to define the direction of the director at the center of the $\pi$-disclination, and one therefore calls these textures ``topological" meaning that there is a loss of continuity at the core.  The topological nature of this excitation makes it very similar to a vortex, for which there is no way to define the phase of the order parameter at the center.  In the case of a vortex, this lack of continuity causes the density to vanish at the core.
For the experimentally relevant spin-1 gases, the density interaction is much stronger than the spin interaction, and it is favorable to fill the core with ferromagneticaly ordered atoms.

Thus, angular momentum is carried by $\pi$ disclinations with ferromagnetic cores.  The spins in the core align perpendicular to the plane in which the nematic directors lie.  In the fast rotating limit, one finds a square lattice of these $\pi$ disclinations with their cores aligned in an antiferromagnetic checkerboard pattern.

The $\pi$ disclination is described by (\ref{rot1}) with $n=m=1/2$.   In a basis with spin projections (1,0,-1), the spinors take the form
\begin{equation}
|n=m=1/2\rangle = \left(\begin{array}{c}
1\\
0\\
b z
\end{array}\right),
\end{equation}
where the real number $b$ is given by minimizing (\ref{var1})
The $\pi$-disclination has a lower energy than the uniform nematic state $(\psi_1,\psi_0,\psi_{-1})=(0,1,0)$ if and only if $\eta>2-2\sqrt{2 c_2}+{\cal O}(c_2)$.  This is a first order (discontinuous) phase transition.

\begin{figure}
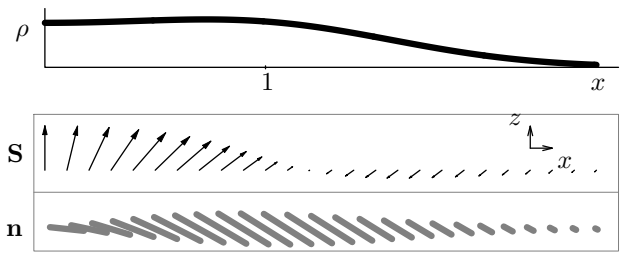

\issmall
{\includegraphics[width=\columnwidth]{slowfigrenum6.lr.epsf}}
{\includegraphics[width=\columnwidth]{slowfigrenum6.eps}}
\caption{Structure of the nematic ring texture (K).  Horizontal axes represent distance from the center of trap along the $x$ axis, measured in units of the trap length.  Top panel shows the density $\rho$.  At each position in the middle panel, 
an arrow is drawn which represents the direction and strength of the local spin.  At each position in the bottom panel, a rod points in the direction of the nematic director $\bf n$, corresponding to the largest eigenvalue of $Q^{(s)}$.  The length of these directors are scaled by the total density, so that their lengths are related to the amount of local nematic order.  To construct the full three-dimensional spin texture, one rotates this picture around the origin, so that the spins near the center look like the skyrmion in figure~\ref{four}, and the nematic order away from the center forms a crown texture.}\label{nemring}
\end{figure}

We find one other azimuthally symmetric texture in the gas with antiferromagnetic interactions. 
This state has the same mathematical structure as (\ref{sym}), except that here $b>0$, while in the ferromagnetic case $b<0$.  As sketched in figure~\ref{nemring}, this texture consists of a nearly uniform nematic ring with a ferromagnetic core.  The spins in the ferromagnetic core bend like the skyrmion in figure~\ref{four}.  The nematic ring has its directors canted slightly from the $\bf \hat z$ axis, forming a crown-shaped texture.  We do not find any composite textures which include this structure.

%has the same shape
%has a `crown' te, where both the spins and the nematic directors point in the $\bf \hat z$ direction.  Closer inspection reveals that the ferromagnetic core has a slight `skyrmion' texture, where as one moves from the center of the cloud, the ferromagnetic order parameter rolls from pointing straight up to having a small component in the x-y plane.  Likewise, the nematic order is not quite uniform, but forms a `crown' texture, where the nematic director is canted radially by some small angle.
%This distortion away from uniform is necessary because in the boundary between the nematic and ferromagnetic regions, the nematic director must locally be perpendicular to the spins.  Far from the origin, the nematic order once again vanishes, and there exists ferromagnetic order in the $-\bf\hat z$ direction, though the density is vanishingly small at these distances.  We do not find any composite textures which include this structure.

\subsection{Composite Textures}\label{composite}
At higher rotation speeds, angular momentum is not carried by single `elemental' textures, but rather by a small collection of these objects.  At very fast rotation speeds one expects to find a regular lattice.   We study these more complicated objects by minimizing (\ref{var1}), sequentially constraining the wavefunction to have the symmetries in (\ref{rot1}) through (\ref{partreflect}).

\begin{figure}
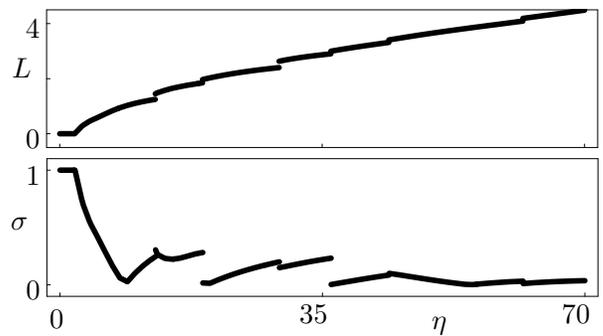

%\psfrag{0}[c]{0}
%\psfrag{35}[c]{35}
%\psfrag{70}[c]{70}
%\psfrag{1}[c]{1}
%\psfrag{4}[c]{4}
%\psfrag{eta}{$\eta$}
%\psfrag{l}{$L$}
%\psfrag{m}{$\sigma$}
%\includegraphics[width=\columnwidth]{fatrib.eps}
\issmall
{\includegraphics[width=234pt]{slowfigrenum8.lr.epsf}}
{\includegraphics[width=234pt]{slowfigrenum8.eps}}
\caption{Angular momentum per particle $L$ and spin polarization per particle $\sigma=\left|\int\!d^2r\,{\bf S(r)}\right|$ of ferromagnetic spin textures as a function of rotation speed, parameterized by $\eta$.  Both are measured in units of $\hbar$.
Note the discontinuities when the ground state symmetry changes.}\label{fatrib}
\end{figure}

\begin{figure}
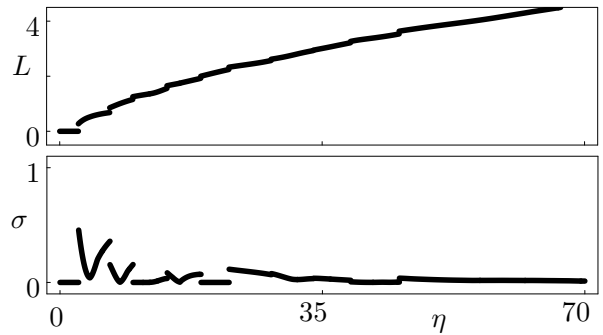

%\psfrag{0}[c]{0}
%\psfrag{35}[c]{35}
%\psfrag{70}[c]{70}
%\psfrag{1}[c]{1}
%\psfrag{4}[c]{4}
%\psfrag{eta}{$\eta$}
%\psfrag{l}{$L$}
%\psfrag{m}{$\sigma$}
%\includegraphics[width=\columnwidth]{natrib.eps}
\issmall
{\includegraphics[width=234pt]{slowfigrenum9.lr.epsf}}
{\includegraphics[width=234pt]{slowfigrenum9.eps}}
\caption{Angular momentum per particle $L$ and spin polarization per particle $\sigma$ of nematic spin textures as a function of rotation speed, parameterized by $\eta$.}\label{natrib}
\end{figure}

\begin{figure}
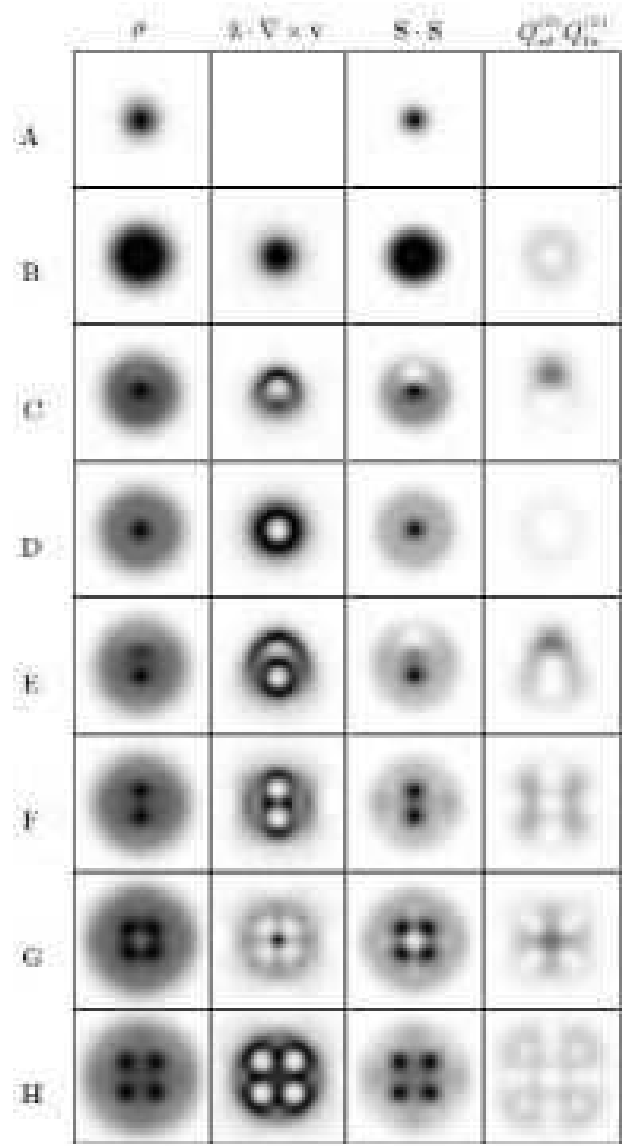

%\psfrag{A}{\small A}
%\psfrag{B}{\small B}
%\psfrag{C}{\small C}
%\psfrag{D}{\small D}
%\psfrag{E}{\small E}
%\psfrag{F}{\small F}
%\psfrag{G}{\small G}
%\psfrag{G}{\small G}
%\psfrag{H}{\small H}
%\psfrag{I}{\small I}
%\psfrag{J}{\small J}
%\psfrag{K}{\small K}
%\psfrag{Lp}{\small L}
%\psfrag{L}{\small L}
%\psfrag{Mp}{\small M}
%\psfrag{M}{\small M}
%\psfrag{N}{\small N}
%\psfrag{O}{\small O}
%\psfrag{Pp}{\small P}
%\psfrag{P}{\small P}
%\psfrag{Q}{\small Q}
%\psfrag{R}{\small R}
%\psfrag{S}{\small S}
%\includegraphics[width=0.842105\columnwidth]{ferrow.eps}
%\includegraphics[width=1.15789\columnwidth]{nemrow.eps}
\issmall
{\includegraphics[width=234pt]{slowfigrenum10.lr.epsf}}
{\includegraphics[width=234pt]{slowfigrenum10.eps}}
\caption{ Spatial structure of spin textures in a gas with ferromagnetic interactions.  From left to right, the columns represent density $\rho$, vorticity $\bf\hat z\cdot \nabla\times v$,  spin density $|S|^2$, and nematic order $\bar Q^{(2)}_{ab} \bar Q^{(2)}_{ba}=\rho^2  Q^{(2)}_{ab}  Q^{(2)}_{ba}$ .  Darker colors corresponds to larger magnitudes.  Although not universally true, $4\pi$ ($8\pi$)-skyrmions tend to show up as black (white) dots in column 2 and white (black) dots in column 3.}\label{struc}
\end{figure}

\begin{figure}
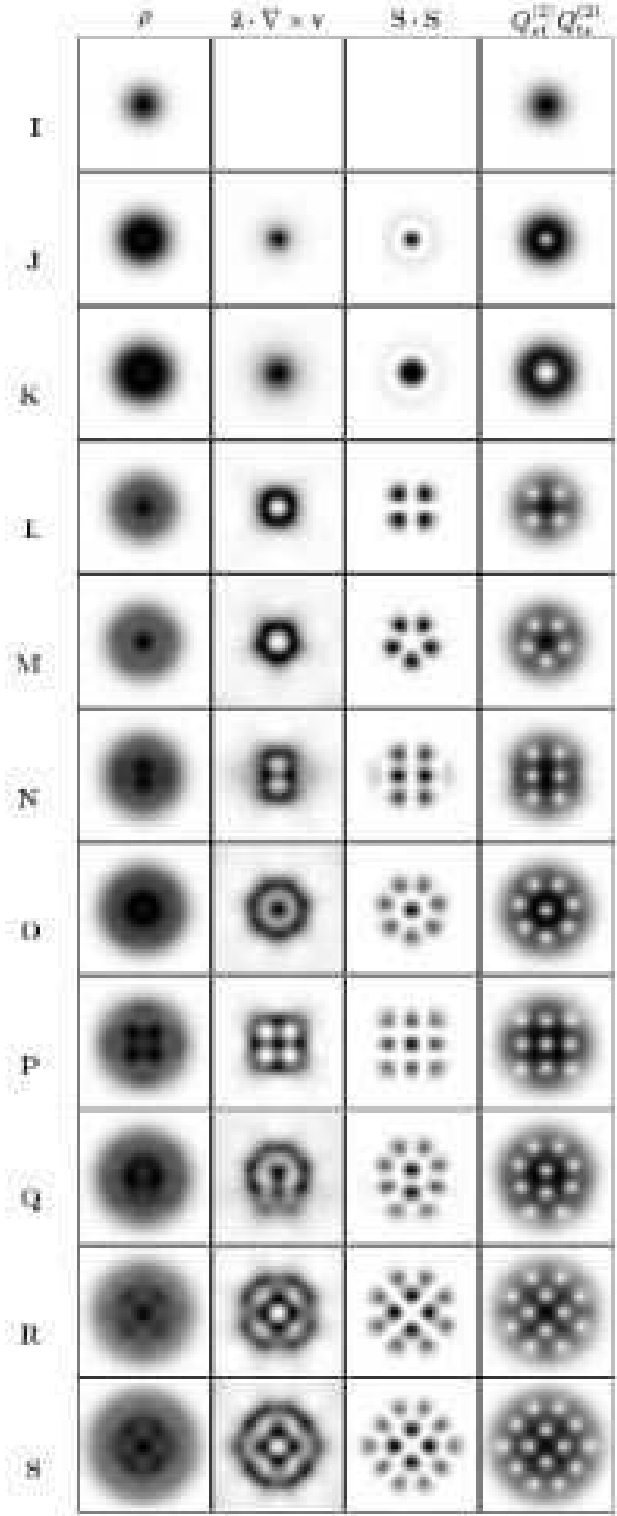

%\psfrag{A}{\small A}
%\psfrag{B}{\small B}
%\psfrag{C}{\small C}
%\psfrag{D}{\small D}
%\psfrag{E}{\small E}
%\psfrag{F}{\small F}
%\psfrag{G}{\small G}
%\psfrag{G}{\small G}
%\psfrag{H}{\small H}
%\psfrag{I}{\small I}
%\psfrag{J}{\small J}
%\psfrag{K}{\small K}
%\psfrag{Lp}{\small L}
%\psfrag{L}{\small L}
%\psfrag{Mp}{\small M}
%\psfrag{M}{\small M}
%\psfrag{N}{\small N}
%\psfrag{O}{\small O}
%\psfrag{Pp}{\small P}
%\psfrag{P}{\small P}
%\psfrag{Q}{\small Q}
%\psfrag{R}{\small R}
%\psfrag{S}{\small S}
%\includegraphics[width=0.842105\columnwidth]{ferrow.eps}
%\includegraphics[width=1.15789\columnwidth]{nemrow.eps}
\issmall
{\includegraphics[width=234pt]{slowfigrenum11.lr.epsf}}
{\includegraphics[width=234pt]{slowfigrenum11.eps}}
\caption{ Spatial structure of spin textures in a gas with antiferromagnetic interactions. 
With the exception of texture K, all black dots in the third column (which also coincide with white dots in the fourth column) can be identified as the cores of $\pi$-disclinations.
}\label{struc2}
\end{figure}

Our numerical results are summarized in figures~\ref{strip} through \ref{struc2}.  Figure~\ref{strip} shows the energy of states with various symmetry properties.  The data is scaled so as to bring out the important features.  There are two parts to the scaling:  (1) ${\cal E}^2$ is plotted rather than just $\cal E$ because in the fast rotating limit one expects ${\cal E}\propto \eta^{1/2}$.  This dependence is made apparent by noting how various terms in the energy scale with the total size of the cloud; $I_2\propto\langle r^2\rangle$ and $I_4\propto1/\langle r^{2}\rangle$.  Minimizing with respect to $\langle r^2\rangle$ then gives ${\cal E}\propto \eta^{1/2}$. (2) We subtract $\beta \eta$ from ${\cal E}^2$, where $\beta$ is an emperically determined constant.  Removing this linear term makes the differences in the curves easier to see.  We use $\beta=-1.7125$ ($-1.8084$) for the ferromagnetic (antiferromagnetic) data. 
%  An inset shows the unscaled data, which clearly gives much less useful information.

The curves in figure~\ref{strip} are labeled by the letters (A) through (S) and can be described as follows:\\
{\bf Ferromagnetic States:}
\noindent{\bf (A)} The uniformly polarized state.
{\bf (B)} The $4\pi$-skyrmion state $|n=1,m=1\rangle$ described in section~\ref{fersec}.  
{\bf (C)}  A texture with a single reflection plane consisting of two elementary structures:
 a $4\pi$ skyrmion near the origin, and a second object at whose center the ferromagnetic order vanishes, to be replace by nematic order.
{\bf  (D)}
The $8\pi$-skyrmion state $|n=2,m=2\rangle$ described in section~\ref{fersec}. 
{\bf (E)} Similar to (C), except the $4\pi$ skyrmion is replace by an $8\pi$ one.  
{\bf (F)}  A texture with two reflection symmetries and a discrete rotation symmetry (with a=2,n=m=1).  Two $8\pi$ skyrmions with their axes canted with respect to one-another lie on one axis, while two {\em weakly} 
nematic regions, where the ferromagnetic order is reduced, lie on the other.
{\bf (G)} A texture with two reflection symmetries and a discrete rotation symmetry (with a=4,n=1,m=0).  At the center lies a nematic region.  Four $8\pi$ skyrmions  with canted axes form a square surrounding the center. 
{\bf (G$^{\prime}$)} as with (G), but the square is distorted into a rhombus, resulting in a lower symmetry (a=2,n=1,m=1).
{\bf (H)} A texture with four reflection axes and a discrete rotation symmetry (a=4,n=1,m=1).  There are four $8\pi$ skyrmions, symmetrically situated with their axes nearly lying in a plane.\\
{\bf Nematic States:}
{\bf (I)} The uniform nematic state.  
{\bf (J)} The $\pi$-disclation with ferromagnetic core described in section~\ref{nemsec}.
{\bf (K)} A nematic ring with a ferromagnetic core described in section~\ref{nemsec}.
{\bf (L)} Four $\pi$-disclinations forming a square.  The feromagnetic cores are aligned antiferromagnetically.  This state has two reflection planes, and a four-fold rotational symmetry (a=4,n=2,m=2).
{\bf (L$^\prime$)} As with (L), but the square is slightly deformed into a rhombus, and the perfect antiferromagnetic alignment of the cores is slightly distorted.  The rotational symmetry is reduced (a=2,n=1,m=1).
{\bf (M)} Five $\pi$ disclinations form a regular pentagon.  The ferromagnetic order at the cores lies completely in the $x$-$y$ plane.  Contains a reflection plane and a five-fold rotational symmetry (a=5,n=2,m=0).
{\bf (M$^\prime$)} As with (M), but the perfect five-fold symmetry is slightly distorted with the pentagon of disclinations stretched along one axis.  This state only has a single reflection plane and no rotational symmetry.
{\bf (N)} Dominated by two stripes of three $\pi$ disclinations organized into a (nearly) square lattice, this texture has a full reflection plane and a partial reflection plane (accross which the spatial but not spin degrees of freedom are reflected.)
{\bf (O)} Eight $\pi$ disclinations:  one at the center, surrounded by seven others. The central core points in the $\hat z$ direction, while the surrounding cores are canted slightly in the $-\hat z$ direction from the $x-y$ plane. Contains one reflection plane and a seven-fold rotational symmetry (a=7,n=3,m=-3).
{\bf (P)} Nine $\pi$ disclinations forming a square lattice with their ferromagnetic cores aligned antiferromagnetically.  This state contains two refection planes and a four-fold rotation symmetry (a=4,n=1/2,m=1/2).
{\bf (P$^\prime$)} As with (P), but the square lattice is deformed towards having a tear-drop shaped set of eight disclinations surrounding the central one.  This state only has a single reflection plane and no rotational symmetries.
{\bf (Q)} A pattern of ten disclinations; two central disclinations surrounded by a distorted oval of eight others.  This state has a single reflection plane.
{\bf (R)} A distorted square lattice, consisting of twelve disclinations. Contains one reflection plane.
{\bf (S)} A distorted square lattice of fourteen disclinations.  Contains two reflection planes, and a discrete rotation symmetry (a=2,n=2,m=2).

Three dimensional animated representations of these states are stored on the epaps archive \cite{epaps}.

From this data one can calculate several observables including: the angular momentum and the degree of spin polarization as a function of $\eta$.  These results are shown in figures~\ref{fatrib} and \ref{natrib}.  The angular momentum is measurable through collective mode experiments \cite{scissor}, while spin polarization could be measured through magnetic susceptability.
In figures~\ref{struc} and \ref{struc2}, we show spatial distributions of the density $\rho$, vorticity $\bf\hat z\cdot\nabla\times v$,  spin density $|S|^2$, and nematic order $Q^{(2)}_{ab} Q^{(2)}_{ba}$ for each of these textures.

\section{Discussion}
\subsection{Alternative classification of skyrmions}
The division of a given texture into elemental skyrmions is not unique, as is illustrated by the two ways of thinking about the $8\pi$ skyrmion in section~\ref{fersec}. 
Above we chose to discuss composite textures in terms of the spin behavior at the density maxima (which is analogous to looking at the $8\pi$ skyrmion from the angle in figure~\ref{eight}).  This is a natural taxonomy scheme in that it reduces the total number of skyrmions which need to be considered, at the cost of introducing $8\pi$ skyrmions.
It is enlightening to also  describe the texture in terms of the spin behavior at the density minima (analogous to the view of the $8\pi$ skyrmion in figure~\ref{doublerot} in terms of four smaller textures).  The advantage of this latter viewpoint is
illustrated by considering texture (H).   At the local minima of the density, the spin points into or out of the page in a manner suggestive of two interpenetrating square lattices.  The textures around each minima can be described as Mermin-Ho skyrmions  (or merons) \cite{mermho} where at the center the spin points in the $\pm\hat z$ direction, and then rolls over to lie in the plane.  In the vicinity of each texture the spins trace out $2\pi$ steradians, covering half of a sphere. 

Under this interpretation the following states have new descriptions:
{\bf (D)} Four skyrmions form a square.  Two of their quantization axes point into the page, two out of the page.
{\bf (E)}  Along the x-axis the density has three minima, and there are correspondingly three elemental textures, two skyrmions and a nematic region. {\bf (F)}  Three skyrmions lie along the x-axis.  {\bf (G)}  Four skyrmions form a square and a nematic region sits at the center.  {\bf (G$^\prime$)} As (G) but the skyrmions form a rectangle.  {\bf (H)}  Two interpenetrating square skyrmion lattices with four $4\pi$ skyrmions pointing into the page and five pointing out.

\subsection{Experimental Consequences}
In this section we address the questions of how to experimentally create and measure these spin textures.  

{\bf Creation:}
One should be able to create these spin textures by the same techniques used to create vortex lattices in scalar condensates \cite{vortexperiments}.  These methods include; `stirring' the cloud with a detuned laser, rotating an ellipsoidal trap, and cooling a rotating cloud through the BEC phase transition.  Several caveats must be kept in mind however: (i) stray magnetic fields and magnetic field gradients must be minimized, and (ii) the large degree if degeneracy in this system combined with experimental randomness may lead to more complicated spin textures than those seen here.  In particular, at fast rotation speeds one would expect to find a domain structure, where different parts of the clouds contain different lattices of skyrmions/disclinations.

\begin{figure}
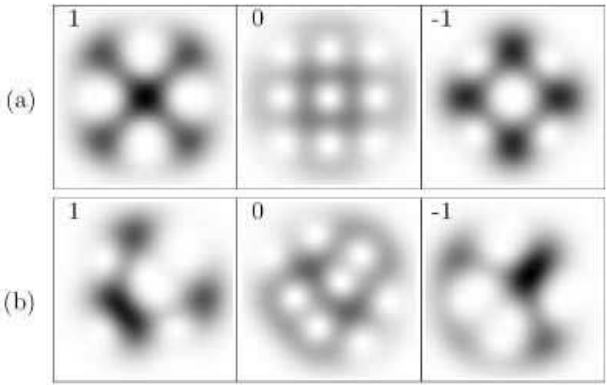

%\psfrag{1}{1}
%\psfrag{-1}{-1}
%\psfrag{0}{0}
%\psfrag{a}[r]{(a)}
%\psfrag{b}[r]{(b)}
%\includegraphics[width=\columnwidth]{components.eps}
\issmall
{\includegraphics[width=234pt]{slowfigrenum12.lr.epsf}}
{\includegraphics[width=234pt]{slowfigrenum12.eps}}
\caption{Densities $|\psi_1|^2$, $|\psi_0|^2$, and $|\psi_{-1}|^2$ of the three components of texture (H), as seen from two different quantization axes.   Darker colors represent higher density.  Panel (a) uses the natural quantization axis where the skyrmion axes are aligned with the $\bf \hat z$ direction, while (b) uses an arbitrary axis.}\label{comp1}
\end{figure}

{\bf Detection:}\label{observe}
We propose directly imaging these textures.
%There are several schemes for imaging the various textures.
%These spin textures can be detected by using various imaging methods.  
First, one turns off the atomic trap, allowing the cloud of atoms to expand.  If a magnetic field gradient is introduced during the expansion, the different components ($\psi_1,\psi_0,\psi_{-1}$) will become spatially separated as in a ``Stern-Gerlach" experiment.  Each of the three components can then be separately imaged. 
This detection method has been used to observe skyrmion textures in the pseudospin-1/2 case at JILA \cite{jilaphaseimprint} and in the spin-1 and 2 case at MIT \cite{ketterleskyrmi}.
 An illustrative density profile for the three components of texture (H) is shown in figure~\ref{comp1}.    Note that the Hamiltonian (\ref{ham}) is invarient under a global rotation of all the spins.  In principle, this means that an experiment may create spin textures with a randomly chosen global orientation.  Figure~\ref{comp1} (a) and (b) shows the components of the same spin texture,  but with the spins uniformly rotated.  The differences between the two figures 
demonstrates how sensitive the component densities are to global spin orientations.  

A second detection scheme makes use of the birefringent properties of a spinor condensate.   As shown in \cite{iaccopo}, light, when detuned from the fundamental transition of the atoms by a frequency which is large compared to the hyperfine structure but small compared to the fine structure,  interacts with the spin textures in a simple manner.  
%**********
%Locally, the index of refraction tensor of the light is proportional to the spin order parameter tensor $Q_{ab}$.  In particular, the ferromagnetic regions of the cloud (where $(Q_{ab}-Q_{ba})/i\neq 0$) are optically active, while the nematic regions are not.  
At these frequencies, the ferromagnetic regions are {\em optically active}, meaning that circularly polarized light aligned with the spins travels at a different speed than the opposite circular polarization.  Consequently, the polarization axis of
linearly polarized light rotates when it passes through a ferromagnetic region.  The angle of rotation is proportional to the projection of the ferromagnetic order along the light propegation direction.  No such rotation occurs when the light passes through a nematic region.  

Using this effect, one can envisage an experimental setup where the sample is imaged with polarized light, but with a crossed polarizer in front of the detector.  Only light which has its polarization axis rotated by passing through a ferromagnetic region will reach the detector.  Thus, one could directly image the ferromagnetic cores of the $\pi$ disclinations found in the antiferromagnetic gas.

\subsection{Quantum Hall physics}
At even higher rotation speeds, quantum fluctuations are expected to melt the regular lattices of spin textures.  The states produced from this melting are highly nontrivial, with strongly-correlated structures reminiscent of the multilayer quantum hall effect \cite{quantum_Hall_papers}.  Exactly how the textures discussed here are connected with the correlated states are a matter of current research.

%It is tempting to conjecture that several of the low angular momentum states found in exact diagonalization calculations are symmetry averaged versions of the textures seen here.

\section{acknowledgement}
I would like to thank Dan Rokhsar, Kareljan Shoutens, and Michael Berry for respective discussions of the symmetries of small clusters of vortices in scalar condensates, reference \cite{reijnders}, and section~\ref{localvort}.
I am indebted to Tin-Lun Ho for his advice, support, and critical comments.
This work was partially supported by NASA Grants NAG8-1441, NAG8-1765, and by NSF Grants
DMR-0109255, DMR-0071630.  

\appendix
\section{Local Vorticity}\label{merminho}
In this appendix we derive equation (\ref{mho}), which relates the curl of the velocity to the spin order parameter.  Introducing a scaled order parameter $\phi_c=\psi_c/\sqrt{n}$, the superfluid velocity is given by
\begin{equation}
{\bf v_s} = \frac{\hbar}{2 i m} (\phi_c^* \nabla \phi_c -\phi_c \nabla \phi_c^*),
\end{equation}
where repeated indices are summed over.  The vorticity is thus given by
\begin{equation}\label{vs}
{\bf \nabla\times v_s} = \frac{\hbar}{i m} (\nabla \phi_c^*\times \nabla \phi_c).
\end{equation}
This expression is related to (\ref{mho}) by noting that 
\begin{eqnarray}
Q_{ab} \nabla Q_{bc}\times \nabla Q_{ca}
&=&\label{q1}\phi_a^* \phi_b \phi_c\phi_a \nabla \phi_b^*\times\nabla\phi^*_c\\
&&\label{q2}+\phi_a^* \phi_b \phi^*_b\phi^*_c \nabla \phi_c\times\nabla\phi_a\\
&&\label{q3}+\phi_a^*\phi_b \phi_c^*\phi_c \nabla\phi_b^*\times\nabla\phi_a\\
&&\label{q4}+\phi_a^*\phi_b \phi_b^*\phi_a \nabla\phi_c\times\nabla\phi_c^*.
\end{eqnarray}
The terms (\ref{q1}) and (\ref{q2}) vanish on account of respectively being antisymmetric in the indices $b,c$ and $a,c$.  Noting that $\phi_a^*\nabla\phi_a=
\nabla(\phi_a^*\phi_a)-\phi_a \nabla \phi_a^*$, one sees that the term (\ref{q3}) is antisymmetric in the indices $a,b$ and therefore also vanishes.  Finally, using $\phi_a^*\phi_a=1$, one finds
\begin{equation}
Q_{ab} \nabla Q_{bc}\times \nabla Q_{ca}=-\nabla \phi_c^* \times \nabla \phi_c,
\end{equation}
which combined with (\ref{vs}) yields (\ref{mho}).  Note that this result is not dependent on the atoms being spin~1, but is completely general.


\begin{thebibliography}{99}
\bibitem{var1} Erich J. Mueller, {\em unpublished}.  The variational calculation can be carried out in the same manner as the spin-1 case which is detailed below.
\bibitem{whyreal}  The imaginary part of $\langle S_\mu S_\nu\rangle$ is related to the spin by the commutation relationships
$[S_\mu,S_\nu] = i \epsilon_{\mu\nu\tau} S_\tau$.
\bibitem{misnomer} Here, and in future discussions of lattices, the designation $4\pi$ skyrmion is a bit of a misnomer, as the spins of each skyrmion only trace out $2\pi$ steradians of solid angle.  These `half-skyrmions' are often referred to as merons.
\end{thebibliography}
\end{document}